\documentstyle[sprocl]{article}

\begin{document}

\title{Gravitational radiation sources for acoustic detectors}

\author{Lee Samuel Finn}

\address{Physics and Astronomy Department, Northwestern University,
2145 Sheridan Road, Evanston IL 60208-3112}

\maketitle\abstracts{}

\section{Experiment, Observation and Astronomy}

Astronomers observe.

Most other sciences permit experiment of some kind, where cause is
manipulated and effect monitored; as astronomers, however, we are
denied this luxury and must learn what we can of the Universe by
eavesdropping on Nature as she murmurs to herself. Denied the luxury
of experiment, we rely on experience guided by intuition
to interpret, and so bring order, to our observations. Our
experience is so far removed from the environment we are studying,
however, that it is a poor guide to our imagination.

An important lesson of astronomy is that the ways we see determine the
things we know. We thought the Universe a different place when, before
Galileo, we saw it only through the naked eye: the real failure of the
Ptolemaic model was only manifest when we saw that the planets had
disks and that Venus passed through a full set of phases. Each
increase in sensitivity, spatial or temporal resolution, or spectral
range has brought with it new puzzles and surprises, and ultimately
new insights and understanding. Progress in astronomy has historically
been associated with new or improved observational perspectives.

Thus, to compensate for the vast difference between our own and
Natures laboratories, we have learned to rely on a multiplicity of
observational perspectives: we view the Universe in as many different
ways as we can in the hope that the multiplicity of distinct
perspectives will lead us to a more reliable view of Nature's
workings.

When completed, the gravitational wave detectors now proposed or under
construction will provide us with a perspective on the Universe
fundamentally different from any we have come to know. With this new
perspective comes the hope that new insights and understandings of
Nature will emerge. The proposed acoustic detectors, with spherical
geometries and operating at millikelvin temperatures, are particularly
well-suited to study gravitational radiation in the 1--3~KHz band. In
this brief report I review some sources of particular interest for
these detectors.

\section{The detector}\label{sec:detector}

For the purpose of describing the sensitivity to astrophysical
gravitational radiation sources of the next generation acoustic
detectors, I assume an array of
TIGA\cite{johnson93a,merkowitz95a,merkowitz96a} ``spheres'' operated
together as a broadband detector. Each element in the array has a
different resonant frequency $f_0$ and a bandwdith $\Delta f_0\simeq
f_0/10$; over this bandwidth the one-sided noise power spectral
density in either $h_+$ or $h_\times$ is constant. The resonant
frequencies and bandwidths of the different array elements are
arranged so that they adjoin with negligible overlap so that broadband
detector composed of these separate elements covers the band from
850~Hz to 2650~Hz with a one-sided noise power spectral density of
approximately $10^{-46}\,{\rm Hz}^{-1}$ in either $h_+$ or
$h_\times$.  Finally, outside of this band I assume that the noise
power is effectively infinite.

\section{Lumpy neutron stars}

Among the more interesting gravitational radiation sources for a high
frequency detector like that described in section \ref{sec:detector}
is a rapidly rotating, non-axisymmetric neutron star.\cite{new95a}
Setting aside pulsar spin-down over the course of an observation as
well as the time-varying Doppler shift arising from Earth's rotation
on its axis and revolution about the Sun, the radiation from such a
lumpy neutron star, rotating about a principle moment of inertia with
rotational period $1/f_{{\rm rot}}$, is
\begin{eqnarray}
h_{+} &=& {4\over r}{1+\cos^2i\over2}\left(2\pi
f_{{\rm rot}}\right)^2\epsilon I_3\cos4\pi f_{{\rm rot}} t\\
h_{\times} &=& -{4\over r}\cos i\left(2\pi
f_{{\rm rot}}\right)^2\epsilon I_3\sin4\pi f_{{\rm rot}} t
\end{eqnarray}
where $I_j$ are the neutron star's principal moments of inertia,
rotation is about the $I_3$ axis, $I_2-I_1=\epsilon I_3$ and $i$ is
the inclination angle between the neutron star angular momentum and
the detector line-of-sight. The mean square signal-to-noise (averaged
over all $i$) in our hypothetical detector for an observation of
duration $T$ is
\begin{equation}
\left<\rho^2\right>
=4.5\times10^4
{T\over{\rm yr}}{10^{-46}\over S_h(f)}
\left(10\,{\rm Kpc}\over r\right)^2
\left(f_{{\rm rot}}\over{\rm KHz}\right)^4
\left({\epsilon\over10^{-6}}{I_3\over10^{45}\,{\rm g cm}^2}\right)^2.
\label{eqn:rho2lumpy}
\end{equation}

In order to interpret the signal-to-noise ratio we can evaluate the
{\em false alarm rate:} the probability that, in a fixed observation,
noise alone will result in a signal-to-noise ratio above a fixed
threshold.
A search for a continuous-wave source in a time series of duration $T$
can achieve a frequency resolution no better than $1/T$. The filtered
detector output in each frequency ``bin'' of width $\Delta f=1/T$ is
independent. The filtered output can be further decomposed into an
in-phase and quadrature-phase component (relative to an arbitrary,
monochromatic reference signal), which are also independent.  The
total signal-to-noise, which is the Pythagorean sum of the
signal-to-noise of the in-phase and quadrature filtered components, is
Rayleigh distributed:
\begin{equation}
P(\rho) = \rho\exp\left(-\rho^2/2\right);
\end{equation}
consequently, the probability that $\rho<\rho_0$ is
\begin{equation}
C(\rho_0) = 1-\exp\left(-\rho_0^2/2\right)
\end{equation}
and the probability that, across all bins in a bandwidth $f$, all the
signal-to-noise ratios are less than $\rho_0$ is
\begin{equation}
P(\rho<\rho_0) = C(\rho_0)^{fT}.
\end{equation}
The false alarm rate ({\em i.e.,} the probability that at least one
bin has $\rho>\rho_0$) is thus $1-C(\rho)^{fT}$. Assuming (as we have
in section \ref{sec:detector}) a detector bandwidth of $1150$~Hz and
an observation period of 1~yr, $\rho>7.5$ corresponds to a false alarm
rate (odds that one or more of the $3.6\times 10^{10}$ ``bins'' are
above threshold owing to noise) of 2.\%, or 1 false alarm every
$50\,{\rm y}^{-1}$.

The likely maximum strength of the neutron star crust
provides an upper bound\cite{pines72a,alpar85a} on $\epsilon$ of
$10^{-6}$. Beyond this upper bound, theory provides no guidance on
$\epsilon$. Observation, on the other hand, does. Gravitational
radiation emission contributes to the spin-down of a rapidly rotating
neutron star; thus, the observed spin-down rate of any millisecond
pulsar limits $\epsilon$ for that pulsar. Observed millisecond pulsars
have very low spin-down rates and correspondingly low upper-bounds on
$\epsilon$: no observed millisecond pulsar admits an $\epsilon$
greater than a few times $10^{-8}$ (Blair, private communication).

Were it known that for some part of the millisecond neutron star
population $\epsilon$ did take on a reasonably large value then one
could use the expected ms~neutron star space density
($n\sim10^{-6}{\rm pc}^{-3}$ {\em in the disk}
\cite{johnston91a,phinney94a}) to estimate the expected number with
signal-to-noise greater than some threshold $\rho_0$. Let's do
that. Suppose that 1\% of ms~neutron stars have
$\epsilon\simeq10^{-7}$ {\em and} rotational frequencies in the range
$100<f_{{\rm rot}}/{\rm Hz}<1000$. Assume that these are distributed
in frequency so that $dn/df\propto f^{-2}$, where $n$ is the number
space density.  Take the galactic disk radius to be 15~Kpc, with Sol
8~Kpc from the galactic center, and the ms~pulsar disk scaleheight $H$
to be 300~pc (the same as for LMXBs). Restricting attention to those
sources with $\rho>7.5$ in a 1~y observation we can expect
\begin{itemize}
\item 27 sources out to the near edge of the galactic disk in the band
$850\,{\rm Hz} < f_{\rm gw} < 1\,{\rm KHz}$;
\item 240 sources in the band
$1\,{\rm KHz} < f_{\rm gw} < 1.8\,{\rm KHz}$;
\item 350 sources throughout the galaxy in the band
$1.8\,{\rm KHz} < f_{\rm gw} < 2\,{\rm KHz}$.
\end{itemize}

A precessing neutron star will also radiate gravitationally in a
manner very similar to that of a non-axisymmetric neutron star
rotating about a principal axis. For the small wobble angles that
might be expected in a realistic source, the radiation amplitudes are
typically much smaller than for the case studied here. The interested
reader may find details in \cite{zimmerman79a,zimmerman80a}.

\section{Coalescing compact binaries}

Coalescing binary neutron star systems are a significant gravitational
wave source for the ground-based inteferometric gravitational wave
detectors like LIGO and VIRGO. These systems radiate most of their
energy at low frequencies, making them less suited as sources for the
proposed acoustic detectors. Nevertheless, it is instructive to
consider how observable compact binary systems are to the proposed
acoustic detectors.

The amplitude signal-to-noise ratio for binary coalescence in an
omni-directional detector is given by\cite{finn93a,finn96a}
\begin{equation}
\rho = 8\left(r_0\over d_L\right)\left({\cal M}\over
1.2\,{\rm M}_\odot\right)^{5/6}\Theta\zeta(f_{{\rm coal}}),
\end{equation}
where $r_0$ is a characteristic distance that depends only on the
source waveform and the detector noise power spectrum,
$d_L$ is the binary's luminosity distance, $\Theta$ depends on the
angular orientation of the binary with respect to the line of sight to
the detector and is of order unity, $\zeta$ describes the fraction of
the detector bandwidth that is filled by radiation from the
inspiraling binary system, and $f_{{\rm coal}}$ is the {\em orbital
frequency} ({\em not} the gravitational wave frequency) where the
inspiral ends and the coalescence begins. For binaries with equal mass
components,\cite{kidder93a}
\begin{equation}
f_{\rm coal} \simeq 710\,{\rm Hz}{2.8\,{\rm M}_\odot\over M},
\end{equation}
where $M$ is the binary system's total mass, and for the detector
described in section \ref{sec:detector},
\begin{eqnarray}
\zeta(f) &=& \left\{
\begin{array}{ll}
0&\mbox{if $f<f_{\min}/2$},\\
{f_{\min}^{-4/3}-(2f)^{-4/3}\over f_{\min}^{-4/3}-f_{\max}^{-4/3}}&
\mbox{if $f_{\min}<f/2<f_{\max}$,}\\
1&\mbox{if $f>f_{\max}/2$}
\end{array}
\right.\\
f_{\min} &=& 850\,{\rm Hz},\\
f_{\max} &=& 2650\,{\rm Hz},\qquad{\rm and}\\
r_0 &=& 28\,{\rm Mpc}\,\left(10^{-46}\,{\rm Hz}\over S_n\right).
\end{eqnarray}
For neutron star binaries ($m_1=m_2=1.4\,{\rm M}_\odot$) $\zeta$ is
approximately $1/2$.

The overlap function $\zeta$ taken together with the relationship
between the system mass and the coalescence frequency $f_{\rm coal}$
shows immediately that the {\em inspiral radiation\/} from binaries
with total mass greater than about $4.7\,{\rm M}_\odot$ will not leave
any imprint on the proposed detectors (however, see the discussion of
black hole formation in section \ref{sec:bh}). Similarly, the value of
$r_0$ for the proposed detector suggests that, in any event, binaries
not much further than the Virgo cluster will be visible. Assuming that
the rate density of neutron star binary
inspiral\cite{narayan91a,phinney91a} is $3\,{\rm y}^{-1}$ at 200~Mpc,
neutron star binary coalescence within 20~Mpc is not expected more
frequently than once every 330~y. The inspiral of ``conventional''
compact binaries is thus not a source of interest for the advanced
proposed acoustic detectors.

What of unconventional binary systems, however? Recent results from
the MACHO collaboration suggest that perhaps 50\% of the galactic halo
mass is in dark objects with $m\sim0.3\,{\rm
M}_\odot$.\cite{turner96Pa} Assuming that
\begin{itemize}
\item a fraction $w\simeq50\%$ of the halo mass
($M_{{\rm halo}}$ greater than or approximately equal to
$6\times10^{11}\,{\rm M}_\odot$) is in MACHOs,
\item all of the MACHOs are mass $m\simeq0.3\,{\rm M}_\odot$ neutron
stars or black holes,
\item a fraction $x\simeq30\%$ of the MACHOs are bound in symmetric
binaries,
\item this binary population was formed at the same time that the
galaxy was formed (age $T\simeq10^{10}\,{\rm y}$), and
\item the MACHO binary population is coalescing today at a steady rate,
\end{itemize}
then the rate of MACHO binary coalescence in the halo is given by
\begin{equation}
\dot{N} = 15\,{\rm y}^{-1} \left(
{x\over 30\%}
{w\over 50\%}
{M_{{\rm halo}}\over6\times10^{11}\,{\rm M}_\odot}
{10^{10}\,{\rm y}\over T}
\right)
\end{equation}
If we take the distance to the typical halo binary to be $50\,{\rm
Kpc}$, then the amplitude signal-to-noise in our antenna array of a
typical halo inspiral is
\begin{equation}
\rho \simeq 510.
\end{equation}
Thus, under these (optimistic!)  assumptions, a TIGA antenna array
would see virtually all halo binary inspirals: 15 strong
($\rho\simeq500!$) events per year.

\section{Black hole formation}\label{sec:bh}

When a binary whose total mass $M$ is greater than the maximum neutron
star mass coalesces the likely outcome is a mass $M$ black hole.  The
time dependence of the spacetime strain owing to the excitation of
fundamental quadrupole mode of the resulting blackhole is, at late
times, given by
\begin{equation}
m(t) = e^{-\pi ft/Q}\sin\left(2\pi ft\right)
\end{equation}
where $Q$ and $f$ are related to the black hole mass and angular
momentum.

In general there are five fundamental quadrupole modes as well as a
succession of ``overtones.'' The fundamental modes are degenerate if
the black hole is non-rotating (Schwarzschild) and are split
otherwise. The overtones are at higher frequency and are also more
strongly damped than the fundamental modes. To give a rough estimate
of the detectability of the gravitational radiation from the late
stages of black hole formation, let's ignore both rotation and
the overtones and assume, for convenience, that black hole
excitation is concentrated in a single mode; then\cite{finn92a}
\begin{eqnarray}
f&\simeq&12.\,{\rm KHz}\left({{\rm M}_\odot\over M}\right), \\
Q&\simeq&2.
\end{eqnarray}

For Schwarzschild black holes, the proposed detector is sensitive to
formation of black holes with mass between 4.5 and 14.1~${\rm
M}_\odot$.  For comparison, the mass of Cygnus X-1 is estimated to be
in the range 5--10~${\rm M}_\odot$, LMC X-3 is about $10\,{\rm
M}_\odot$, and A0620-00 is about $3.8\,{\rm M}_\odot$; so, the
formation of astrophysical black holes in our antenna array's ``mass
range'' is reasonable.

Assuming a quadrupole radiation pattern for the black hole we can
relate the ``efficiency'' of the black hole formation (ratio of the
total energy radiated to the black hole mass) to the rms (averaged
over all orientations of the black hole with respect to the detector
line-of-sight) strain induced in each element of our antenna array:
\begin{equation}
h_{{\rm rms}}(t) = 2\sqrt{\epsilon}{M\over r}m(t)
\end{equation}
Numerical modeling\cite{anninos93a} of Schwarzschild black hole
head-on collisions give efficiencies of $10^{-4}$. For collisions of
rotating black holes with significant orbital angular momentum the
efficiencies might be significantly higher --- perhaps 1\% or greater.

The mean square amplitude signal-to-noise ratio of Schwarzschild black
hole formation in our hypothetical detector is thus given
by\cite{finn92a}
\begin{eqnarray}
\overline{\rho^2} &\simeq& 34.
\left(\epsilon\over10^{-4}\right)
\left(20\,{\rm Mpc}\over r\right)^2
\left(M\over13\,{\rm M}_\odot\right)^3
\left(10^{-46}\,{\rm Hz}^{-1}\over S_h(f)\right).
\end{eqnarray}
Even under the pessimistic assumption that there is no increase in
efficiency for the coalescence of orbiting black holes our antenna
array is sensitive to, we can expect to observe a typical $10\,{\rm
M}_\odot$ black hole formation event with a signal-to-noise of several
out to the Virgo cluster.

The rate of black hole formation is entirely uncertain; however, most
astrophysicists see no reason why the same mechanisms that make
neutron stars do not also make black holes at approximately the same
rate. By our present understanding of formation mechanisms, this is
not a high rate even at the distance of the Virgo cluster; so,
evidence of black hole formation at any measurable rate would require
a significant change in our understanding of stellar evolution.

\section{Supernovae}

Theoretical models of stellar core collapse, and the corresponding
gravitational wave luminosity, have a long and checkered history:
estimates of the gravitational wave luminosity have at different
times ranged over more than four orders of magnitude. It is not
simply the luminosity that is unknown: the waveforms themselves are
also uncertain, leading to a further difficulty in estimating the
detectability of this source. Nevertheless, it is still possible to
evaluate what is required of stellar core collapse in order that it be
observable in our hypothetical detector.

Suppose that the waveform from supernovae is given by
\begin{eqnarray}
h_{+} &=& {2{\rm M}_\odot\over r}\alpha f_{+} m(t)\\
h_{\times} &=& {2{\rm M}_\odot\over r}\beta f_{\times} m(t)
\end{eqnarray}
$\alpha$ and $\beta$ are constants, $f_{+}$ and $f_\times$ are
functions of the relative orientation of the source with respect to
the detector, and $m(t)$ is some function of time which we leave
undetermined for now. The power radiated into each polarization mode
is given by
\begin{eqnarray}
\dot{E}_{+} &=& \alpha^2\left<f^2_+\right>{\rm M}_\odot^2|\dot{m}|^2\\
\dot{E}_{\times} &=& \beta^2\left<f^2_{\times}\right>{\rm
M}_\odot^2|\dot{m}|^2,
\end{eqnarray}
where $<>$ signifies an {\em average\/} over the sphere.

Now assume that equal power is radiated into the two polarization
modes. Then we can write $\alpha$ and $\beta$ in terms of a single
parameter $\epsilon$ as
\begin{eqnarray}
\alpha^2 &=& {\epsilon\over2{\rm M}_\odot\left<f^2_+\right>
\int dt\,|\dot{m}|^2}\\
\beta^2 &=& {\epsilon\over2{\rm M}_\odot\left<f^2_{\times}\right>
\int dt\,|\dot{m}|^2}.
\end{eqnarray}
In terms of $\epsilon$ the power radiated into the $+$ and $\times$
polarization states is thus
\begin{equation}
\dot{E}_+ = \dot{E}_\times =
{\epsilon{\rm M}_\odot|\dot{m}|^2 \over 2\int dt |\dot{m}|^2}.
\end{equation}

Now let us return to consider the time dependence of the
waveform $m(t)$. Note that
\begin{equation}
\int dt\,|\dot{m}(t)|^2 = \int df\,(2\pi f)^2|\tilde{m}(f)|^2,
\end{equation}
where $\tilde{m}$ is the Fourier transform of $m$. Assume that there
is equal radiated power in equal bandwidths out to a frequency $f_0$;
then $|\tilde{m}|^2f^2$ is constant for $f<f_0$ and zero for
$f>f_0$. The mean-square signal-to-noise ratio (averaged over the
source orientation relative to the detector) in the detector described
in section \ref{sec:detector} is then
\begin{equation}
\left<\rho^2\right> = {\epsilon{\rm M}_\odot\over r^2}{1\over 2\pi^2S_n f_0}
\left({1\over f'_{\min}}-{1\over f'_{\max}}\right)
\end{equation}
where
\begin{eqnarray}
f'_{\min} &=& \max\left[f_{\min},\min(f_0,f_{\min})\right]\\
f'_{\max} &=& \max\left[f_{\min},\min(f_0,f_{\max})\right].
\end{eqnarray}
For the detector described in section \ref{sec:detector}, $f_{\min}$ is
850~Hz and $f_{\max}$ is 2650~Hz. If we assume that $f_{\min}<f_0\simeq1\,{\rm
KHz}<f_{\max}$, then
\begin{equation}
\left<\rho^2\right> \simeq
420.{\epsilon\over10^{-4}}
\left({100\,{\rm Kpc}\over r}\right)^2
{10^{-46}\,{\rm Hz}^{-1}\over S_n}
{10^3\,{\rm Hz}\over f_0}
{850\,{\rm Hz}\over f_{\min}}
{\left(1-{f_{\min}/f_0}\right)\over0.15}
\end{equation}

Current calculations (which, bear in mind, are still unable to
successfully describe the supernova explosion) suggest $\epsilon$ in
the range $10^{-9}$--$10^{-8}{\rm M}_\odot$, with peak power in the
200--300~Hz band.\cite{finn91a,monchmeyer91a,burrows96a} Thus, without
a very optimistic efficiency and a substantially wider signal
bandwidth, we can't expect to be able to observe supernovae much
beyond our own galaxy, and certainly not out to the Virgo cluster.

\section{Conclusions}

There are several interesting opportunities for an array of
millikelvin acoustic gravitational antennae covering the 1--3~KHz
frequency band. In particular, we can expect that the radiation from
population of rapidly rotating neutron stars with oblateness on order
1/10 that allowed by the crustal breaking strength would be observable
throughout the galaxy in a one year observation. Local supernovae,
neutron star binary inspiral, or black hole formation, while
infrequent, would be a serendipitous radiation source from our own
galaxy or, perhaps, as far as the Virgo cluster. Finally, a very
speculative source --- coalescence of a population of compact MACHOs
in the galactic halo --- would be observable with large
signal-to-noise ratio.

\section*{Acknowledgments}

It is a pleasure to thank the meeting organizers for their efforts in
conducting what, by all measures, was a very successful and enjoyable
meeting. It is a pleasure to acknowledge the generous support of the
Alfred P. Sloan Foundation and the National Science Foundation (PHY
95-03084).

% bibliographystyle is already set at top
%\bibliography{phyjabb,references,preprints,keys}

\end{document}